\documentclass[12pt,a4paper,twoside,final,notitlepage, leqno]{article}
\usepackage[english]{babel}
\usepackage[T1]{fontenc}  
\usepackage{graphicx}
\usepackage{amsmath,amsthm,epsfig,amsfonts,bbm}  
\newcommand{\pequationdeb}{$$ \left\{ \begin{minipage}[c]{130mm}}
\newcommand{\pequationfin}{\end{minipage}
                           \right. $$}

\newcommand{\moneq}{\vspace*{-6pt} \begin{equation} \displaystyle } 
\newcommand{\moneqstar}{\vspace*{-6pt} \begin{equation*} \displaystyle } 
\newcommand{\monendstar}{\vspace*{-6pt} \end{equation*}   }
\newcommand{\monend}{\vspace*{-6pt} \end{equation}   }
\newcommand{\beq}     {\begin{equation}}
\newcommand{\enq}     {\end{equation}}
\newcommand{\be}    {\begin{enumerate}}
\newcommand{\ee}    {\end{enumerate}}

\newcommand{\Bb}

\def\R{{\rm I}\! {\rm R}}
 
\def\tvi {\vrule  height 10pt depth 5pt width 0pt}
\def\tv  {\tvi \vrule}
\def\tvg {\tv ~~}
\def\tvd {~~ \tv}
\def \na{ \noalign {\hrule}  }
\def\hcr {\hfill & \cr}
\def\br {\break}
\newcommand{\C}[0]{\mathbb{C}}
\def\section*#1{}
\def\resume{\if@twocolumn
\section*{R\'esum\'e}
\else \small
\quotation{\bf \it R\'esum\'e \rule[1mm]{1.5mm}{0.2mm}\vspace{0pt}}
\fi}
\def\endresume{\if@twocolumn\else\endquotation\fi}
\def\abstract{\if@twocolumn
\noindent\section*{{\bf Abstract}}
\else \small
\quotation{\noindent \bf {Abstract.} \rule[1mm]{1.5mm}{0.2mm}\vspace{0pt}}
\fi}
\def\endabstract{\if@twocolumn\else\endquotation\fi}

\usepackage{fancyhdr}
\renewcommand{\headrulewidth}{0pt}
\begin{document}

%%%%%%%%%%%%%%%%%%%%%%%%%%%%%%        07 fevrier 2012 
\fancypagestyle{plain}{ \fancyfoot{} \renewcommand{\footrulewidth}{0pt}}
\fancypagestyle{plain}{ \fancyhead{} \renewcommand{\headrulewidth}{0pt}} 
%%%%%%%%%%%%%%%%%%%%%%%%%%%%%%        fin 07 fevrier 2012   

\title {\bf \LARGE    ~  \vspace{.5 cm} ~\\  On quantum models for \\ 
opinion and voting intention   polls
\\ ~  }

\author { { \large  Fran\c{c}ois Dubois~$^{ab}$}    \\ ~\\ 
{\it  \small $^a$   Conservatoire National des Arts et M\'etiers,  }  \\      
{\it \small  Department of Mathematics, Paris, France.} \\     
{\it \small  $^b$  Association Fran\c caise de Science des Syst\`emes} \\ 
{\it \small francois.dubois@math.u-psud.fr}  } 

\date { 10  september 2013~\protect\footnote{~This contribution has been presented
to  the seventh international Quantum Interaction conference, 
%%% Leicester, UK,  on 25th July 2013.
and is published in 
{\it Quantum Interaction~- 7th International Conference, QI2013}, 
Leicester, UK,   25-27 July 2013,
Editors  H. Atmanspacher, E. Haven, K. Kitto, D. Raine,
Lecture Notes  in Computer Science, Springer, 
volume 8369,  pages~286-295, 2014. Edition May 2023.}}

  \bigskip    \bigskip  

\maketitle

\begin{abstract} 
In this contribution, we construct a connection between two quantum voting models
presented previously. 
We propose to try to determine the result of a vote from associated given opinion polls.
We introduce a density operator relative to the family 
of all candidates to a particular   election. 
From  an hypothesis of proportionality between  a  family of coefficients 
which characterize the density matrix and 
 the probabilities  of vote  for all the candidates, 
we propose a numerical method for the entire determination of the density operator. 
This approach is a direct consequence of the Perron-Frobenius theorem
for irreductible positive matrices. 
We apply our algorithm  to synthetic data and to operational 
results  issued from the French presidential election of April 2012. 
 $ $ \\[4mm]
   {\bf Keywords}: density matrix, Perron-Frobenius theorem, French presidential election.   
 $ $ \\[4mm]
   {\bf AMS classification}: 65F15, 81Q99, 91C99. 
%%%%%%%%%%%%%    35L40   First-order hyperbolic systems
%%%%%%%%%%%%%    58J45   Hyperbolic equations, in Global analysis, analysis on manifolds
%%%%%%%%%%%%%    81Q99   Quantum theory 
%%%%%%%%%%%%%    91C99   Social and behavioral sciences: general topics
%%%%%%%%%%%%%    65F15   Eigenvalues, eigenvectors
%
\end{abstract}

%%%%%%%%%%%%                         31 aout 2013                      %%%%%%%%%%%%%%%%  
\bigskip \bigskip  \newpage \noindent {\bf \large 1) \quad  Introduction}  
%%%%%%%%%%%%                         31 aout 2013                      %%%%%%%%%%%%%%%%  
 
%%%%%%%%%%%%%%%%%%%%%%%%%%%%%%        07 fevrier 2012 
\fancyfoot[C]{\oldstylenums{\thepage}}
%%%%%%%%%%%%%%%%%%%%%%%%%%%%%%        fin 07 fevrier 2012   

  \noindent $\bullet$ \quad  
Electoral periods are favorable to opinion polls. 
We keep in mind that 
opinion polls  are intrinsically complex
(see {\it e.g.}  Gallup \cite{Ga44}) %%% or  Till\'e \cite {Ti01}) 
%%% or the introduction of Bar-Hen and Chiche \cite{BC09})
and give an approximates picture of a possible social reality. 
They are traditionnally of two types: popularity polls  for various 
outstanding political personnalities %%%   on one hand 
and  voting intention   polls    when a list of candidates is known. 
%
%%%%%%     We remark that in the first case, a grid of appreciation 
%%%%%%    is given by the questionnaire, typically of the type 
%%%%%%    \quad  ``very  good''  $\, \succ \,$ 
%%%%%%    ``good''        $\, \succ \,$  
%%%%%%    ``no opinion''      $\, \succ \,$  
%%%%%%    ``bad''      $\, \succ \,$  
%%%%%%    ``very bad''.
%
%%%%%%   \smallskip \noindent $\bullet$ \quad  
We have   two different informations and to construct a link between them is not an easy task. 
In particular, the determination of the voting intentions is a quasi intractable  problem!
Predictions of votes classically use   of so-called ``voting functions''.
Voting functions  have been developed for the
prediction of presidential elections in the United States. 
They are based on correlations 
between economical parameters, popularity polls and other technical parameters. 
%
%%%   of the outgoing president, 
%%%   popularity and predictions of elections have been developed. 
%  
We refer to  Abramowitz \cite {Ab88}, Lewis-Beck \cite {LB91},  
Campbell  \cite{Ca92} and  Lafay \cite{La97}.
%%%%%%%%%      and the survey paper proposed by  Auberger \cite{Au04}. 

  \smallskip \noindent $\bullet$ \quad  
We do not detail here  the mathematical difficulties associated with the question 
of voting when the number of candidates is greater than  three 
\cite  { Ar51, Bo1781, Co1785}. 
They conduct to present-day researches like range voting,  
  independently proposed by Balinski and Laraki \cite{BL07a, BL07b} 
% at Ecole Polytechnique (Paris) 
and  by Rivest and Smith   \cite {Sm2k, RS07}. 
%%%%  at  the ``Center of   Range Voting'' (Stony Brook, New York). 
%
 It is composed by two steps: grading and ranking. In the  grading step, 
all the candidates are evaluated by all the electors. 
This first step is quite analogous to a popularity  investigations and 
we will merge the two notions in this contribution. 
The second step of range voting is a  majority ranking;
it consists of a successive extraction of medians. 
%%%   Note that this theory is  applied for wine testing, 
%%%   as described in  Peynaud and  Blouin \cite  {PB06}.

  \smallskip \noindent $\bullet$ \quad  
In this contribution, we adopt   quantum modelling 
(see {\it e.g.} Bitbol {\it et al} \cite{Bi09} for an introduction), 
in the spirit of authors like   
 Khrennikov and Haven \cite{HK07, HK13}, La Mura and  Swiatczak \cite{LMS07} 
and Zorn and  Smith \cite{ZS11} concerning voting processes.   
The fact of  considering  quantum modelling induces a specific vision of probabilities. 
We refer {\it e.g.} to the classical  treatise  on quantum mechanics 
of  Cohen-Tannoudji {\it et al.} \cite{CDL77}, 
to the so-called contextual objectivity proposed by Grangier  \cite{Gr02},   
%% to the  approach of Mugur-Sch\"achter  \cite{MMS08},   
or to the elementary introduction 
proposed by Busemeyer and  Trueblood \cite  {BT09}
in the context of statistical inference.

  \smallskip \noindent $\bullet$ \quad  
This contribution is organized as follows:
we recall in Sections~1 and~2 two quantum models for the vote developed previously \cite{Du08, Du09}
and a first tentative \cite{Du12} to connect these two models (Section~4). In Section~5, we develop
the main idea of this paper. We construct a link between opinion polls and voting.
This idea is tested numerically in Sections~6 and~7 for synthetic data and a ``real life''
 election.

%%%%%%%%%%%%                         31 aout 2013                      %%%%%%%%%%%%%%%%  
 \bigskip \bigskip  \newpage \noindent {\bf \large 2) \quad  A fundamental  elementary model }  
%%%%%%%%%%%%                         31 aout 2013                      %%%%%%%%%%%%%%%%  
 
\noindent $\bullet$ \quad  
In a  first tentative \cite{Du08}, we have proposed to introduce an 
Hilbert space  $\, V_\Gamma \,$ formally generated by the candidates 
$ \, \gamma_j \in \Gamma . \, $ In this space, a candidate $ \, \gamma_j \,$
is represented by a unitary vector $ \,   |  \, \gamma_j \! >  \,$ 
and this family of $n$ vectors is supposed to be orthogonal. 
Then an elector $ \, \ell \,$ can be decomposed 
in the space   $\, V_\Gamma \,$ of candidates according to
\moneq  \label{decompo-ell} 
\displaystyle 
 |  \, \ell \! > \,\,=\, \,
\sum_{j=1}^n  \, \theta_j  \,\,  | \, \gamma_j \! >  \, .     
\monend 
The vector $ \,  |  \, \ell \! > \,  \in V_\Gamma \,$ is supposed  to be 
a unitary vector to fix the ideas.  
According to Born's rule, the probability for a given elector 
$\, \ell \, $ to give his  voice to the particular candidate
$ \, \gamma_j \, $ is equal to $ \,  |  \,   \theta_j  \,  | ^2 . \, $
The violence of the quantum measure is clearly visible  with this example: 
the opinions of an elector $ \, \ell \,$ never coincidate with the
program of any candidate. But with a voting system where an elector  has to choice 
only one candidate among $n$, his social opinion is {\it reduced} 
to the one of a  particular candidate.

%%%%%%%%%%%%                         31 aout 2013                      %%%%%%%%%%%%%%%%  
 \bigskip \bigskip \noindent {\bf \large 3) \quad  A quantum model for range voting }  
%%%%%%%%%%%%                         31 aout 2013                      %%%%%%%%%%%%%%%%  
 
\noindent $\bullet$ \quad  
Our  second model \cite{Du09} is adapted to  the grading step 
of range voting \cite{BL07a,Sm2k}.  
We consider a grid $\rm G$ of $m$ types of opinions as one of the two following ones.  
We have $ \, m=5 \,$ for the first grid (\ref{grille-5}) and  
$ \, m=3 \,$ for the second one  (\ref{grille-3}):   
\moneq  \label{grille-5}
 ++  \,\, \,\succ \, \,\, 
 +  \,\,\, \succ \,\,\,
 0  \,\,\, \succ \,\,\, 
 -  \,\,\, \succ \,\,\,
 -- 
\monend 
\vskip -.7  cm
\moneq  \label{grille-3} 
 +  \,\,\, \succ \,\,\,
 0  \,\,\, \succ \,\,\, 
 -   \,. 
\monend 
These ordered grids are typically  used for popularity polls.  
%%%%%%        \cite {If12a, If12b, Ip12, Ip09avr12}.
%
We assume also that a ranking grid like 
(\ref{grille-5}) or (\ref{grille-3}) is a basic tool  
to represent a social state of the opinion. 
%%%%%%%%  
We introduce a specific grading space $ \, W_{\rm G} \,$ of political appreciations 
 associated with a grading
family $\rm G$. The space  $ \,  W_{\rm G} \,$ is formally generated by the $m$ orthogonal
vectors  $ \,  |  \, \zeta_i  \! > \,$ relative to the opinions. 
Then we suppose that the candidates $ \, \gamma_j \,$ are now decomposed 
by each elector %% $\, \ell \,$ 
on the basis   $ \,  |  \, \zeta_i \! >  \,$ for $ \, 1 \leq i \leq m$:
\moneq  \label{decompo-gamma} 
\displaystyle 
   |  \, \gamma_j \! > \,=\, \sum_{i=1}^m \, \alpha_j^i \, \, 
   |  \, \zeta_i \! > \, , \quad \gamma_j \in \Gamma \, , \quad 1 \leq j \leq n \, .  
 \monend 
Moreover the vector  $ \,  |  \, \gamma_j \! >  \,$ in   (\ref{decompo-gamma})
is supposed to be by a unitary:  
\moneq  \label{gamma-unitaire} 
\displaystyle 
 \sum_{i=1}^m\,   |  \,  \alpha_j^i  \,  | ^2  \,\, = \,\, 1 \, ,
\quad \gamma_j \in \Gamma \, , \quad 1 \leq j \leq n \, . 
 \monend 
With this notation,  %%  (where we have omitted the index $\ell$), 
the probability for a given elector to appreciate  a candidate $ \, \gamma_j \,$
with  an opinion $ \, \zeta_i \,$ is simply a consequence of the Born  rule.
The mean statistical
expectation   of a given opinion $\, \zeta_i \,$ for a candidate  $ \, \gamma_j \,$ is equal to  
 $ \,  |  \,   \alpha_j^i \,  | ^2  \, $  on one hand and is given by  
the popularity polls  $ \,  S_{j \, i}  \,  $  on the other hand. Consequently, 
\moneqstar %%%   \label{alpha-deux-egal-s} 
  |  \,  \alpha_j^i  \,  | ^2 \,=\,  S_{j \, i} 
\,, \quad \gamma_j \in \Gamma \,, \,\, \zeta_i \in {\rm G} \,, \quad  1 \leq j \leq n \,, \quad 
1 \leq i \leq m . 
 \monendstar   
%    

%%%%%%%%%%%%                         31 aout 2013                      %%%%%%%%%%%%%%%%  
 \bigskip \bigskip \noindent {\bf \large 4) \quad  A first link between the two previous models }  
%%%%%%%%%%%%                         31 aout 2013                      %%%%%%%%%%%%%%%%  

\noindent $\bullet$ \quad  
In \cite{Du12}, we have proposed a  first link between the two previous models.  
 We   simplify   the approach  (\ref{decompo-ell})
and suppose that there exists some equivalent candidate 
$ \,  |  \, \xi \! > \, \in V_\Gamma \,$ such that the voting intention
%%%   $ \, \beta_j \,$ 
for each particular candidate $ \, \gamma_j \in  \Gamma  \,$
is equal to $\,  | <\xi \,,\, \gamma_j  >  |^2 $. 
%
%%%    \moneq  \label{equivalent-candidate} 
%%%   | <\xi \,,\, \gamma >  |^2 \,=\,  \beta_j \,, \quad \forall  \, \gamma \in  \Gamma 
%%%  \,; \quad  |  \, \xi \! > \, \equiv \, \sum_{\gamma \in \Gamma}  |  \, \gamma \! >
%%%  \,  <\gamma \,,\, \xi  > \, \,\, \in V_j \, . 
%%%   \monend 
% 
We interpret   the relation (\ref{decompo-gamma}) in the following way: for each
candidate $ \, \gamma_j \in \Gamma, \,$
there exists a political decomposition  $ \, A  \, |  \, \gamma_j \! > \, \in W_{\rm G} \,$
in terms of the grid G.   %%%%%%%%%%       and we have 
%
%%%%%%%%%%     \moneq  \label{opinion-gamma} 
%%%%%%%%%%     \displaystyle  
%%%%%%%%%%     A  \, |  \, \gamma \! >  \,=\, \sum_{i=1} \, \alpha_j^i \, \, 
%%%%%%%%%%        |  \, \zeta \! > \, , \quad \gamma \in \Gamma \, .  
%%%%%%%%%%      \monend 
%    
By linearity, we   construct in this way  a linear operator 
$\, A :\, V_\Gamma  \longrightarrow W_{\rm G} \,$ between two different Hilbert spaces. 
Preliminary results have been presented, in the context of the 2012 French presidential
election.

%%%%%%%%%%%%           31 aout, 7 septembre  2013             %%%%%%%%%%%%%%%%  
%%  \bigskip \bigskip \noindent {\bf \large 5) \quad  Tentative of unification in the space of opinions  }  
 \bigskip \bigskip \noindent {\bf \large 5) \quad  From opinion polls to the prediction of the vote }  
%%%%%%%%%%%%           31 aout, 7 septembre  2013             %%%%%%%%%%%%%%%%  

\noindent $\bullet$ \quad  
In the  space $ \, W_{\rm G} \,$ of political appreciations described in Section~3 of this
contribution, the opinion polls allow through the relation 
(\ref{decompo-gamma}) to determine some knowledge about each candidate 
$\, \gamma_j \in \Gamma \,$ in %%  an orthonormal basis $\,  |  \, \zeta_i \! > \,$ of 
 the space  $ \, W_{\rm G}$. We suppose that each candidate is represented by
a unitary vector and the relation (\ref{gamma-unitaire}) still holds. 
The question is now to evaluate the probability for an arbitrary elector to vote for the 
various candidates. 

\smallskip \noindent $\bullet$ \quad  
We denote by $ \, \Pi_j \equiv  | \, \gamma_j  \! > < \! \gamma_j \, | \,$ 
the orthogonal projector onto the direction
of the state  $\, | \, \gamma_j  \! > $. Then 
we introduce a density matrix $ \, \rho \,$ associated to a statistical
representation of the voting population:
\moneq  \label{matrice-densite} 
\displaystyle   
\rho \,=\,  \sum_{j=1}^n \alpha_j \,   \Pi_j  \, \equiv \, 
\sum_{j=1}^n \alpha_j \,   | \, \gamma_j  \! > < \! \gamma_j \, | \, .
 \monend 
% 
%%%%%%%%%%%%%%%%%%
%%%%%%%   Le lien avec les positive-operator valued measure (povm)
%%%%%%%   reste a faire...   Il est hors sujet ici... 
%%%%%%%%%%%%%%%%%%
%
It is classical that $ \, {\rm tr}  \rho = \sum_{j=1}^n \alpha_j \,$ 
and if   $ \,  \alpha_j \geq 0 \,$ for each index $j$,  the auto-adjoint operator 
 $ \, \rho \,$  is non-negative:
\moneqstar  
\displaystyle   
<  \! \rho \, \zeta \,,\, \zeta \! >  \, \, \geq \, \, 0 \,, \qquad 
\forall \zeta \in  W_{\rm G}   \, .
 \monendstar 
It is then natural to search the coefficients $ \, \alpha_j \,$ such that 
\moneq  \label{coefficients}   \left \{ \begin{array}{l}
\displaystyle   
\alpha_j \,\geq \, 0 \,, \qquad  1 \leq j \leq n \\\displaystyle   
\sum_{j=1}^n \alpha_j \, = \, 1 \, .   
 \end{array} \right.   \monend
In these conditions, the esperance of election $ \, < \gamma_j > \,$ 
of  the candidate $ \, \gamma_j \,$  is given through the relation
\moneq  \label{esperance-election}   
< \gamma_j > \,=\, {\rm tr} \big( \rho \, \Pi_j \big) \,, \quad 1 \leq j \leq n \, . 
 \monend
We have the following calculus:

\smallskip \noindent $ \displaystyle 
{\rm tr} \big( \rho \, \Pi_j \big) \,=\, \sum_{k=1}^n \, 
< \zeta_k \,, \, \rho \, \Pi_j \, \zeta_k > 
\,=\,  \sum_{k=1}^n  \sum_{\ell=1}^n \, 
< \zeta_k \,, \, \alpha_\ell \, \gamma_\ell > \, < \gamma_\ell \,, \, \gamma_j > 
 \, < \gamma_j \,, \, \zeta_k >  $ 

\smallskip \noindent $ \displaystyle \qquad \qquad  = \, 
 \sum_{\ell=1}^n \,    \alpha_\ell \,  < \gamma_\ell \,, \, \gamma_j > \, 
 \sum_{k=1}^n \, < \zeta_k \,, \, \gamma_\ell > \,< \gamma_j \,, \, \zeta_k >  $

\smallskip \noindent $ \displaystyle \qquad \qquad  = \, 
 \sum_{\ell=1}^n \,    \alpha_\ell \,  < \gamma_\ell \,, \, \gamma_j > \, 
 \sum_{k=1}^n \, < \gamma_j \,, \, \zeta_k >  \,  < \zeta_k \,, \, \gamma_\ell >$ 

\smallskip \noindent $ \displaystyle \qquad \qquad  = \, 
 \sum_{\ell=1}^n \,    \alpha_\ell \,  < \gamma_\ell \,, \, \gamma_j > \, 
< \gamma_j  \,, \, \gamma_\ell > \,= \, 
 \sum_{\ell=1}^n \,    \alpha_\ell \,  \mid < \gamma_\ell \,, \, \gamma_j > \mid ^2   \, . $ 

\smallskip \noindent
We introduce the matrix $A$  composed by the squares of the scalar products of the vectors of candidates:
\moneq  \label{matrice-A} 
A_{j \, \ell} \,=\, \, \mid < \gamma_j \,, \, \gamma_\ell > \mid ^2 \, , 
\quad 1 \leq  j,\,  \ell \leq  n \, . 
 \monend
Then the previous calculs establishes that 
\moneq  \label{valeur-moyenne} 
< \gamma_j > \,=\, \sum_{\ell=1}^n \,  A_{j \, \ell} \, \, \alpha_\ell  \, . 
 \monend

\smallskip \noindent $\bullet$ \quad  
It is interesting to imagine a link between the esperance of election
$ \, < \gamma_j > \,$ of the candidate $ \, \gamma_j \,$ and the coefficient
$ \, \alpha_j \,$ of the density matrix introduced in (\ref{matrice-densite}). 
In general they differ. In the following, 
we focus our attention to the particular case 
 where these two quantities are proportional, {\it id est} 
\moneq  \label{proportionnel} 
\exists \, \lambda \in \C \,, \, \quad \forall \, 
j = 1, \, 2, \, \dots \, , \, n  \,, \, \quad
  < \gamma_j > \,=\, \lambda \,  \alpha_j  \, . 
 \monend
Because both $ \, < \gamma_j > \,$  and  $\, \alpha_j \,$ are positive, the coefficient  
$ \, \lambda \,$ must be a positive number. Moreover, due to (\ref{valeur-moyenne}), 
the relation (\ref{proportionnel}) express that the non-null vector $ \, \alpha 
\in \R^n \,$ composed by the coefficients $ \, \alpha_j \,$ is an eigenvector 
of the matrix $A$. Then,  due to the hypothesis  (\ref{coefficients}), we have  
$ \, \alpha_j \geq 0 \,$ and this eigenvector has non-negative components. 
If we suppose that the matrix $A$ is irreductible (see {\it e.g.} in the book of
Meyer \cite{Me2k} or Serre  \cite{Se02}), 
the Perron-Frobenius theorem states that there exists a {\bf unique} eigenvalue
(equal to the spectral radius of the matrix $A$) such that the 
corresponding eigenvector has all non-negative components. Moreover, all the components
of this eigenvector are strictly positive. 
In other words, if the matrix $A$ defined in (\ref{matrice-A}) is irreductible and if the 
hypothesis of proportionality (\ref{proportionnel}) is satisfied, the coefficients
$ \, \alpha_j \,$ of the density matrix are, due to the second relation of (\ref{coefficients}), 
 completely defined.  %%%   by the proportionality hypothesis   (\ref{proportionnel}). 
In the following, we propose to determine the coefficients 
$ \, \alpha_j \,$ of the density matrix (\ref{matrice-densite}) 
and satisfying the conditions (\ref{coefficients}) 
as proportional to the positive eigenvector of the matrix 
$A$ defined by (\ref{matrice-A}).

\smallskip \noindent $\bullet$ \quad  
The above model is not completely satisfactory for the following reason. The 
underlying order associated to the grading family $\, G \,$ has not been taken 
into account. To fix the ideas, we suppose that each grade  $ \, \nu_i \,$ is
associated to a %% positive 
number  $ \, \sigma_i \,$ such that 
\moneq  \label{numerisation} 
\sigma_1 \, > \, \sigma_2    \, > \, \cdots   \, > \, \sigma_m  \, .  \monend
%% \sigma_1 \, > \, \sigma_2    \, > \, \cdots   \, > \, \sigma_m  \geq 0 \, .  \monend
%
We introduce a ``popularity operator''  $ \, P_j \,$ associated 
to the $j$th candidate $ \, \gamma_j$:
\moneq  \label{operateur-popularite} 
P_j \, \equiv  \, \sum_{i=1}^m  \sigma_i \, \,  
 |  < \! \gamma_j \,, \, \zeta_i \! > |^2 \, 
|  \, \zeta_i   \! >  < \! \zeta_i \, |  \, .  \monend
We can determine without difficulty the mean value
of the operator $ \, P_j \,$ for the density configuration 
$ \, \rho \,$ defined in  (\ref{matrice-densite}):

\smallskip \noindent $ \displaystyle 
{\rm tr} \big( \rho \, P_j \big) \,=\, \sum_{k=1}^m \, 
< \zeta_k \,, \, \rho \, P_j \, \zeta_k > $ 

\smallskip \noindent $ \displaystyle \qquad \qquad  = \, 
\sum_{k=1}^m  \sum_{\ell=1}^n \, 
< \zeta_k \,, \, \alpha_\ell \, \gamma_\ell > \, < \gamma_\ell \,, \, 
 \sum_{i=1}^m  \sigma_i \, \,  
 |  < \! \gamma_j \,, \, \zeta_i \! > |^2 \, 
|  \, \zeta_i   \! >  < \! \zeta_i \, , \, \zeta_k >  $ 

\smallskip \noindent $ \displaystyle \qquad \qquad  = \, 
\sum_{i=1}^m  \sum_{\ell=1}^n \,  \sigma_i \,  \alpha_\ell \, 
< \! \zeta_i \, , \,  \gamma_\ell   \! >  \, 
 < \gamma_\ell \,, \, \zeta_i   \! > \, 
|  < \! \gamma_j \,, \, \zeta_i \! > |^2 \,  $ 

\smallskip \noindent $ \displaystyle \qquad \qquad  = \, 
\sum_{i=1}^m  \sum_{\ell=1}^n \,  \sigma_i \,  \alpha_\ell \, 
|  < \! \gamma_\ell \,, \, \zeta_i \! > |^2 \, 
|  < \! \gamma_j \,, \, \zeta_i \! > |^2 \,.  $

\smallskip \noindent 
In other words, if we set 
\moneq  \label{matrice-B} 
B_{j \, \ell} \, \equiv  \, \sum_{i=1}^m \,  \sigma_i \, 
|  < \! \gamma_j \,, \, \zeta_i \! > |^2 \, \,
|  < \! \gamma_\ell \,, \, \zeta_i \! > |^2 \, \,,   \monend
we have:
\moneq  \label{moyenne-popularite} 
< P_j > \,\equiv \, {\rm tr} \big( \rho \, P_j \big) \,=\, \
 \sum_{\ell=1}^n \, B_{j \, \ell} \, \, \alpha_\ell  \, .   \monend

\smallskip \noindent $\bullet$ \quad  
We use  a positive parameter $ \, t \,$ and search the coefficients 
$\, \alpha_j \,$ in such a way that the mean value of the candidate 
$\gamma_j$ with some ``upwinding'' associated to its popularity is 
proportional to the above coefficients. In other words, 
due to (\ref{valeur-moyenne}) and (\ref{moyenne-popularite}), the mean 
value $ \,  < \gamma_j > \, + \,  t < P_j > \, $ takes the algebraic form
\moneq  \label{moyenne-decentree}  
< \gamma_j > \, + \,  t < P_j > \,=\, 
 \sum_{\ell=1}^n \, \big( A_{j \, \ell} + t \, B_{j \, \ell} \big)  \, \alpha_\ell  \, .   \monend
Under the condition that all the coefficients 
$ \,  A_{j \, \ell} + t \, B_{j \, \ell} \,$ are positive, {\it id est} that the 
parameter $t$ is small enough, we compute the coefficients 
$ \, \alpha_\ell \,$ with the help of the Perron-Frobenius
theorem as presented previously. 
%

%%%%%%%%%%%%           8 septembre  2013             %%%%%%%%%%%%%%%%  
 \bigskip \bigskip \noindent {\bf \large 6) \quad  A first numerical test case }  
%%%%%%%%%%%%           8 septembre  2013             %%%%%%%%%%%%%%%%  

\noindent $\bullet$ \quad
Our first model uses synthetic data. We suppose that we have three candidates
($n=3$) and two ($m=2$) levels of ``political''  appreciation. %%  of the candidates.
We suppose that 
\moneq  \label{synthetic-data}   \left \{ \begin{array}{l} 
\displaystyle  | \gamma_1 \! >  \,= \, \cos \Big( {\pi\over6} \Big) \, | \zeta_1 \! > 
\,+\, \sin \Big( {\pi\over6} \Big) \, | \zeta_2 \! >  \\
\vspace {-.4cm}  \\
\displaystyle  | \gamma_2 \! >  \,= \, \cos \Big( {\pi\over4} \Big) \, | \zeta_1 \! > 
\,+\, \sin \Big( {\pi\over4} \Big) \, | \zeta_2 \! >  \\
\vspace {-.4cm}  \\
\displaystyle  | \gamma_3 \! >  \,= \, \cos \Big( {\pi\over3} \Big) \, | \zeta_1 \! > 
\,+\, \sin \Big( {\pi\over3} \Big) \, | \zeta_2 \! >  \, . 
 \end{array} \right.   \monend
With the choice $ \, \sigma_1 = 1 $ and $ \, \sigma_2 = 0 \, $ in a way  
suggested at the relation  (\ref{numerisation}), we can simulate  numerically 
the process presented in the Section~5. The results are presented in  
Figure~1. When the variable $t$ is increasing, the first candidate
has a better score, due to his best results in the grading evaluation 
 (\ref{synthetic-data}). 
 
%%%%%%%%%%%%%%%%%%%%%%%%%%%%%%%%%     Figure 1      %%%%%%%%%%%%%%%%%%%%%%%%%%%%%%  
\smallskip   \smallskip                      
\centerline {  {\includegraphics[width=.75 \textwidth]   {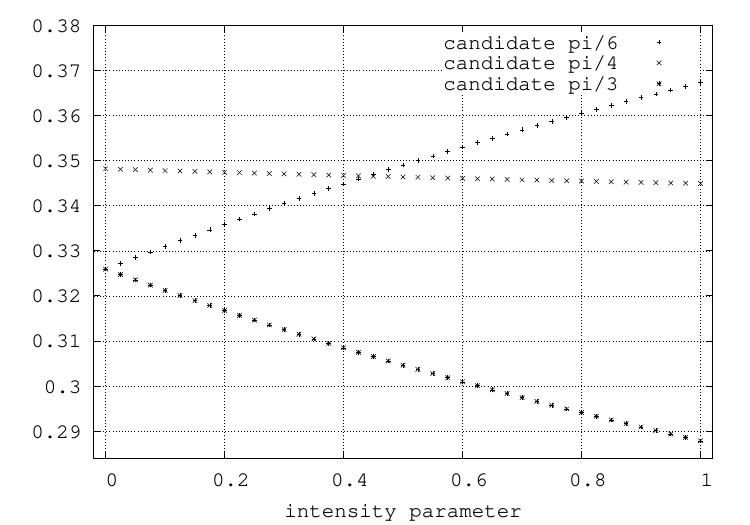}} }  

\smallskip \noindent  {\bf Figure 1}. \quad  Result of the vote obtained by a quantum model
from the opinion poll, with synthetic data proposed in (\ref{synthetic-data}). 

\smallskip \smallskip 
%%%%%%%%%%%%%%%%%%%%%%%%%%%%%%%%%%%%%%%%%%%%%%%%%%%%%%%%%%%%%%%%%%%%%%%%%%%%%%%%%%%%

%

%%%%%%%%%%%%           8 septembre  2013             %%%%%%%%%%%%%%%%  
 \bigskip \bigskip \noindent {\bf \large 7) \quad  Test of the method with real  data }  
%%%%%%%%%%%%           8 septembre  2013             %%%%%%%%%%%%%%%%  

\noindent $\bullet$ \quad 
We have also used data coming from the ``first tour'' of the French presidential election of 
April~2012. 
%  The family of data presented   has been obtained in April~2012. 
Popularity data  \cite{Ip09avr12}
and result of voting intentions  \cite{Ip10avr12}
are displayed in Table~3. The names of the principal candidates 
to the French presidential election 
  are proposed in alphabetic order with the following  abbreviations:
``Ba'' for  Fran\c cois Bayrou,  
``Ho'' for  Fran\c cois Hollande, 
``Jo'' for  Eva Joly, 
``LP'' for  Marine Le Pen, 
``M\'e'' for  Jean-Luc M\'elanchon and 
``Sa'' for  Nicolas Sarkozy. 
In  Table~3, we have also reported the result of the election 
of 22 April 2012. 
 
%%%%%%%%%%%%%%%%%%%%       ``Table 03''   avril   2012                %%%%%%%%%%%%%%%%%%%%%%%%%%%%%% 
%%%%%%%%%%%%%%%%%%%%       sources popularite ipsos publie 9 avril 2012         %%%%%%%%%%%%%%%% 
%%%%%%%%%%%%%%%%%%%%       sources sondage vote ipsos publie 10 avril 2012      %%%%%%%%%%%%%%%% 
   \bigskip   \bigskip 
\setbox20=\hbox{ $\,\,$ }
\setbox30=\hbox{ $  + +  $   }
\setbox40=\hbox{ $  +   $  } 
\setbox50=\hbox{  ~ 0  } 
\setbox60=\hbox{ $  -    $    } 
\setbox70=\hbox{ $ - -   $ } 
\setbox80=\hbox{ voting } 
\setbox90=\hbox{ result }  %%% 22 April 2012
\setbox21=\hbox{ Ba }
\setbox22=\hbox{ Ho } 
\setbox23=\hbox{ Jo } 
\setbox24=\hbox{ LP } 
\setbox25=\hbox{ M\'e }
\setbox26=\hbox{ Sa } 
%%         .55 .14 .31 ;    % Bayrou   .125 ;    % Bayrou  
\setbox41=\hbox{ .56  }  %%% Bayrou  +  
\setbox51=\hbox{ .07  }  %%% Bayrou 0   
\setbox61=\hbox{ .37  }  %%% Bayrou  -    
\setbox81=\hbox{ .095  }  %%% Bayrou intention de vote 
\setbox91=\hbox{ .091  }  %%% Bayrou vote 22 avril 
%%       .52 .08 .40 ;    % Hollande    .30  ;    % Hollande    
\setbox42=\hbox{ .57  }  %%% Hollande +  
\setbox52=\hbox{ .03  }  %%% Hollande 0   
\setbox62=\hbox{ .40  }  %%% Hollande -    
\setbox82=\hbox{ .285 }  %%% Hollande intention de vote   
\setbox92=\hbox{ .286 }  %%% Hollande vote 22 avril 
%%       .29 .13 .58 ;    % Joly       .03  ;    % Joly    
\setbox43=\hbox{ .35  }  %%% Joly +  
\setbox53=\hbox{ .10  }  %%% Joly 0   
\setbox63=\hbox{ .55  }  %%% Joly -      
\setbox83=\hbox{ .015 }  %%% Joly intention de vote 
\setbox93=\hbox{ .023 }  %%% Joly vote 22 avril  
%%       .28 .06 .66 ] ;  % Le Pen    .175 ] ;  % Le Pen   
\setbox44=\hbox{ .26  }  %%% Le Pen +  
\setbox54=\hbox{ .05  }  %%% Le Pen 0   
\setbox64=\hbox{ .69  }  %%% Le Pen -      
\setbox84=\hbox{ .15  }  %%% Le Pen  intention de vote 
\setbox94=\hbox{ .179 }  %%% Le Pen  vote 22 avril    
%%%      .38 .20 .42 ;    % Melanchon   .085 ;    % Melanchon   
\setbox45=\hbox{ .47  }  %%% Melanchon +  
\setbox55=\hbox{ .10  }  %%% Melanchon 0   
\setbox65=\hbox{ .43  }  %%% Melanchon -    
\setbox85=\hbox{ .145 }  %%% Melanchon  intention de vote 
\setbox95=\hbox{ .111 }  %%% Melanchon  vote 22 avril      
%%       .33 .0  .67 ;    % Sarkozy    .25  ;    % Sarkozy   
\setbox46=\hbox{ .49  }  %%% Sarkozy +  
\setbox56=\hbox{ .05  }  %%% Sarkozy 0   
\setbox66=\hbox{ .46  }  %%% Sarkozy -    
\setbox86=\hbox{ .29  }  %%% Sarkozy  intention de vote 
\setbox96=\hbox{ .272 }  %%% Sarkozy  vote 22 avril      
\setbox44=\vbox{\offinterlineskip  \halign { 
&\tvg#& # &\tvg#&  #  &\tvg#& #  &\tvg#&  #&\tvd#&   #&\tvg#&  #&\tvd#&   #&\tvg#&  #&\tvd#\cr   
\na&   \box20  &&  \box40  &&  \box50  && \box 60 && $\!\!$&&  \box80 && $\!\!$&&  \box90  \hcr  
\na&   \box21  &&  \box41  &&  \box51  && \box 61 && $\!\!$&&  \box81 && $\!\!$&&  \box91  \hcr  
\na&   \box22  &&  \box42  &&  \box52  && \box 62 && $\!\!$&&  \box82 && $\!\!$&&  \box92  \hcr   
\na&   \box23  &&  \box43  &&  \box53  && \box 63 && $\!\!$&&  \box83 && $\!\!$&&  \box93  \hcr   
\na&   \box24  &&  \box44  &&  \box54  && \box 64 && $\!\!$&&  \box84 && $\!\!$&&  \box94  \hcr   
\na&   \box25  &&  \box45  &&  \box55  && \box 65 && $\!\!$&&  \box85 && $\!\!$&&  \box95  \hcr   
\na&   \box26  &&  \box46  &&  \box56  && \box 66 && $\!\!$&&  \box86 && $\!\!$&&  \box96  
\hfill   \hcr  \na}   }  \centerline{\box44  }
\smallskip \centerline {  {\bf  Table  1}.   \quad 
 Popularity, sounding polls and result,   april  2012 \cite{Ip09avr12,Ip10avr12}. } 
  \bigskip    

%%%%%%%%%%%%%%%%%%%%%%%%%%%%%%%%%     Figure 2      %%%%%%%%%%%%%%%%%%%%%%%%%%%%%%  
\smallskip   \smallskip                      
\centerline {  {\includegraphics[width=.75 \textwidth]   {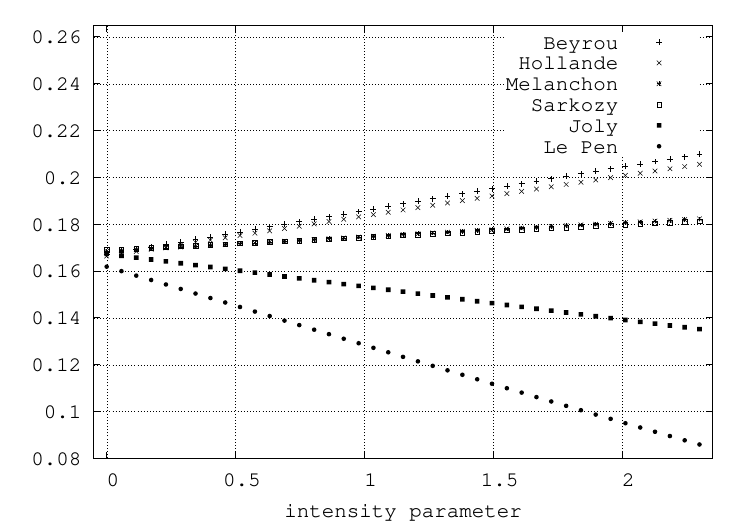}} }  

\smallskip \noindent  {\bf Figure 2}. \quad  Result of the vote obtained by a quantum model
from the opinion poll. Data issued from the April 2012 French presidential election. 

\smallskip \smallskip 
%%%%%%%%%%%%%%%%%%%%%%%%%%%%%%%%%%%%%%%%%%%%%%%%%%%%%%%%%%%%%%%%%%%%%%%%%%%%%%%%%%%%

\smallskip \noindent $\bullet$ \quad 
This test case corresponds to $n=6$ and $m=3$. 
The numerical data relative to  the relation  (\ref{numerisation}) are chosen such that 
 $ \, \sigma_1 = 1 $,  $ \, \sigma_2 = 0 $, and $ \, \sigma_3 = -1  $.
Then the above Perron-Frobenius methodology is available up to 
$ \, t = 2.2 $. The numerical result are presented in Figure~2. It reflects 
some big tendances of the real election. But the correlation between the popularity 
and the result is not always satisfied, as shown clearly by comparison between
our simulation in Figure~2 and 
the result of the election shown in the last column of Table~1.

%%%%%%%%%%%%           8 septembre  2013             %%%%%%%%%%%%%%%%  
 \bigskip \bigskip \noindent {\bf \large 8) \quad  Conclusion  }  
%%%%%%%%%%%%           8 septembre  2013             %%%%%%%%%%%%%%%%  

\noindent $\bullet$ \quad 
In this contribution, we have used a given quantum model for range voting in the context of opinion 
polls. From these data, we have proposed a quantum metodology for predicting the vote. 
We introduce a density operator associated to the candidates. 
The mathematical key point is the determination of a positive eigenvector for 
a real matrix with non-negative coefficients. 
Our results are encouraging, even if the confrontation   %% comparison
to real life data shows explicitly that other parameters have to be taken into account.

 \bigskip

%%  \newpage 
%%%%%%%%%%%%%%%%%%%%%%%%%%%%%%%%%%%%%%%%%%%%%%%%%%%%%%%%%%%%%%%%%%%%%%%%%%%%%%%%%%%%%
%%%%%%%%%%%%%%%%%%%%%%%%%%%%%%%%%%%%%%%%%%%%%%%%%%%%%%%%%%%%%%%%%%%%%%%%%%%%%%%%%%%%%
\bigskip \bigskip  \bigskip % \bigskip  % \newpage 

\medskip

\end{document}